\documentclass[12pt]{article}
\def\be{\begin{eqnarray}}
\def\ee{\end{eqnarray}}
\def\bq{\begin{equation}}
\def\eq{\end{equation}}
\def\ben{\begin{enumerate}}\def\een{\end{enumerate}}

\def\roughly#1{\mathrel{\raise.3ex\hbox{$#1$\kern-.75em%
\lower1ex\hbox{$\sim$}}}}

\begin{document}
\begin{titlepage}

\hfill FTUV  05-1125


 \vspace{1.5cm}
\begin{center}
\ \\
{\bf\LARGE Are monopoles hiding in monopolium?}
\\
\vspace{0.7cm} {\bf\large Vicente Vento} \vskip 0.7cm

{\it  Departamento de F\'{\i}sica Te\'orica and Instituto de
F\'{\i}sica Corpuscular}

{\it Universidad de Valencia - Consejo Superior de Investigaciones
Cient\'{\i}ficas}

{\it 46100 Burjassot (Val\`encia), Spain, }

{\small Email: Vicente.Vento@uv.es}

\end{center}
\vskip 1cm \centerline{\bf Abstract}

Dirac showed that the existence of one magnetic pole in the universe
could offer an explanation of the discrete nature of the electric
charge. Magnetic poles appear naturally in most Grand Unified
Theories. Their discovery would be of greatest importance for
particle physics and cosmology. The intense experimental search
carried thus far has not met with success. I propose a Universe with
magnetic poles which are not observed free because they hide in
deeply bound monopole--anti-monopole states named monopolium. I
study the feasibility of this proposal and establish signatures for
confirming my scenario.

\vspace{2cm}

\noindent Pacs: 14.80.Hv, 95.30.Cq, 98.70.-f, 98.80.-k

\noindent Keywords: nucleosynthesis, monopoles, monopolium

\end{titlepage}

\section{Introduction}

The theoretical justification for the existence of classical
magnetic poles, hereafter called monopoles, is that they add
symmetry to Maxwell's equations \cite{jackson} and explain charge
quantization \cite{dirac1}. Dirac showed that the mere existence of
a monopole in the universe could offer an explanation of the
discrete nature of the electric charge. His analysis leads to the so
called Dirac Quantization Condition (DQC),

\bq \frac{e g}{\hbar c} = \frac{N}{2} \;, \mbox{  N = 1,2,...}\; ,
\eq

\noindent where $e$ is the electron charge and $g$ the monopole
charge. Note that if quarks were asymptotic states the minimum
monopole charge would be three times larger.

The origin of monopoles, and therefore their properties, is diverse.
In Dirac's formulation monopoles are assumed to exist as point-like
particles and quantum mechanical consistency conditions lead to
Eq.(1), establishing the value of their magnetic charge. However,
their mass, $m$, is a parameter of the theory, limited only by
classical reasonings to be $m > 2 $ GeV \cite{book}. In non-Abelian
gauge theories monopoles arise as topologically stable solutions
through spontaneous breaking via the Kibble mechanism \cite{kibble}.
They are allowed by most Grand Unified Theory (GUT) models, have
finite size and come out extremely massive $m > 10^{16}$ GeV.
Furthermore, there are also models based on other mechanisms with
masses between those two extremes \cite{book,giacomelli,rujula}.

The discovery of monopoles would be of greatest importance not only
for particle physics but for cosmology as well. Therefore monopoles
and their experimental detection have been a subject of much study
since many believe in Dirac's statement\cite{dirac1}\\

{\sl "...one would be surprised if Nature had made no use of
it [the monopole]."}\\

At present, despite intense experimental search, there is no
evidence of their existence
\cite{book,giacomelli,review,experiment,mulhearn}. This state of
affairs has led me to investigate a possible mechanism by which
monopoles could exist and still be undetectable by present
experiments.

Despite the fact that monopoles symmetrize  in form Maxwell's
equations there is a numerical asymmetry arising from the DQC,
namely that the basic magnetic charge is much larger than the
smallest electric charge. This led Dirac himself in his 1931 paper
\cite{dirac1} to state,\\

{\sl "... the attractive force between two one quantum poles of
opposite sign is $(\frac{137}{2})^2 \approx 4692\frac{1}{4}$ time
that between the electron and the proton. This very large force
may perhaps account for why the monopoles have never been
separated."} \\

This latter statement sets the philosophy of my scenario: at some
early stage in the expansion of the Universe, monopoles and their
antiparticles were created; at that time, the dynamics was such
that, soon thereafter, most of them paired up to form
monopole-anti-monopole bound states called monopolium. With time,
monopolia become more and more bound until their constituents
annihilate. However, the mean life of monopolium is sufficiently
long to allow monopolia to exist even today in measurable
abundances. Thus today, most of the existing monopoles, appear
confined in deeply bound states \cite{zeldovich}. Monopolia have
produced observable signatures during their formation period, a
remnant isotropic radiation, and provide us with direct evidence of
their present existence in the form of Ultra High Energy Cosmic Rays
(UHECR) and localized low frequency radiation from within the core
of the clusters of galaxies\cite{hill,sigl} .

\section{Hidden monopoles}

I proceed to describe the proposed scenario. Mine is a
phenomenological approach. I envisage a scenario, which is
realized by means of a few assumptions, in which monopoles are not
observable as free states at present, that satisfies all
experiments and leads to new observations which can sustain it.
What cosmological models realize the proposed scenario is for the
present investigation of no relevance.

At some early stage in the expansion of the Universe monopoles and
their antiparticles were created by a mechanism which is free from
the standard monopole problem \cite{kolb}. No precise mechanism
for their creation is advocated, therefore the mass is not fixed
and is left as a parameter to be fitted by consistency
requirements. Moreover, (anti)monopoles existed in the Universe,
at that time, at the level of abundance compatible with known
phenomenological and experimental upper bounds
\cite{ahlen,parker,adams}. Most of the (anti)monopoles bind in
pairs during nucleosynthesis, due to the strong magnetic forces,
to form monopolium. This fundamental hypothesis of my scheme is
realized physically by the following condition,

\bq r_{capture} << \lambda.\label{inequality} \eq
Here $\lambda$ is the mean free path of the monopoles in the hot
plasma,

\bq \lambda \sim \frac{1}{\sigma \rho_{ch}}\; , \eq

\noindent where $\sigma$ is the cross section for the scattering of
the monopole with charged particles \cite{book,zeldovich}

\bq \sigma \sim 2 \, \frac{m c^2}{k T} \;\; {\mbox nanobarns}\; ,
\eq

\noindent $\rho_{ch}$ the density of charged particles and $m$ is
the mass of the monopole; and $r_{capture}$ is the capture distance
of the monopole \cite{book,zeldovich},

\bq r_{capture} \sim \frac{g^2}{k T} \; .\eq

Furthermore, I describe the monopoles from the monopolium formation
era up to the present days by point like Dirac monopoles. It is
reasonable to do so since the discussion is largely independent of
the detailed structure of the monopoles because it depends only on
global properties, i.e., magnetic charge, mass and cosmological
abundances.

In the rest of the paper I proceed to show that my assumptions
lead to a picture consistent with data.

I describe monopolium as a Bohr atom, with reduced mass $m/2$ and a
strong magnetic, instead of a weak electric, coupling. Its binding
energy is

\bq E \sim \left(\frac{1}{8 \alpha}\right)^2 \;\frac{m c^2}{n^2}\;
,\eq

\noindent where $\alpha = \frac{1}{137}$ is the fine structure
constant of $QED$ and $n$ the principal quantum number. This
equation and those that follow are to be considered only for large
principal quantum number ($n > 50$). For low values of $n$ the
annihilation mechanism becomes dominant.

The approximate size of the system is given by

\bq r \; \sim \; <r>_{n,0}\; \sim\; \frac{12 \hbar}{m c}\; \alpha
\; n^2 \; . \eq

\noindent To calculate the mean life I distinguish two processes, i)
the cascading process, dominated by dipole radiation \cite{hill},
which I apply as

\bq \tau_{dipole} \sim \frac{ 2 m^2 c a_n^3}{\hbar^2} \sim 2\;
(12)^3 \;\frac{\hbar}{mc^2}\; \alpha ^5 \;n_i^6 \; ,\eq

\noindent where $n_i$ is the principal quantum number associated
with the initial bound state which will be very large $n_i \sim
10^9$; ii) the annihilation process, which due to the magnitude of
$g$ is highly non perturbative and which I next estimate. Looking at
the two photon decay process I see \cite{dirac2,wheeler}

\bq \tau_{annihilation} < < \tau_{2 \gamma} \sim 2\; (4)^5
\frac{\hbar}{m c^2} \;\alpha^5 \; n_f^3 \; ,\eq

\noindent where $n_f$ is the largest principal quantum number
associated with a state at which annihilation is still efficient.
Since the monopole and anti-monopole only annihilate efficiently
when there is a considerable probability of being on top of each
other and this only happens for $ n < 50$, $n_f << n_i$. Thus the
annihilation mean life is small compared with the cascading time and
can be disregarded in the time scale analysis.

The previous equations can be summarized in terms of the binding
energy of the initial bound state and the mean life as,

\be
E_b (eV) \,  r_b (\mbox{\AA}) & \sim & 5. \;  10^4 eV \mbox{\AA} \\
n_i &\sim & 9. \; 10^4 [E_b(eV)^{1/4} \tau(sec)^{1/4}] \\
m c^2 (eV) & \sim & 3. \; 10^7 [E_b(eV)^{3/2}  \tau(sec)^{1/2}]\; eV
\; . \label{binding}\ee

\noindent Here $E_b$ and $r_b$ are respectively the binding energy
and the radius of the initial bound state.

Using Eqs. (\ref{inequality}) through (\ref{binding}) my scenario
can be constructed. I assume that capture takes place for a binding
energy slightly higher, to avoid thermal dissociation, than $kT = 1$
MeV,

\bq E_b > 1 \mbox{ MeV}\; . \eq
This temperature is not related to the scale for production of
monopoles but to that at which monopolium, the bound state, is
formed from already existing monopoles \cite{hill}.

\noindent From these equations monopolium is a tightly bound
system

\bq r_b < 0.05 \mbox{\AA} < r_{capture} \sim 0.3 \mbox{\AA}\; .\eq

The solution of the remaining equations require a recurrent process
since they are all intertwined. Consistency has been achieved for
the following values.

I obtain for the mean life of monopolium, $\tau$, defined from
$\tau_{dipole}$,

\bq  \tau_U/ \tau \sim 100 \;,\eq

\noindent where  $\tau_U$ is the age of the Universe.

Capture happens at a very outer shell,

\bq n_i \sim 5. \; 10^{9}, \eq

\noindent and the mass of the monopole comes out,

\bq m c^2 \sim 5. \; 10^{15} \mbox{ GeV}\; , \eq

\noindent which justifies the non relativistic treatment for large n
and is surprisingly very close to the mass of the GUT monopole.

Finally I obtain my crucial assumption satisfied as

\bq  r_{capture} \sim 0.3\; \mbox{\AA} << \lambda \sim 5.6 \;
\mbox{\AA} \; ,\label{inequality1}\eq

\noindent where I have used that for $kT \sim 1$ MeV the density of
charged particles is \cite{kolb}

\bq \rho_{ch} \sim 1.8 \; 10^{21} \mbox{ particles/} cm^3 \;.
 \eq

It was pointed out that the drag force felt by the monopole in the
plasma reduces dramatically the mean life of the state
\cite{blanco}. This effect is not active in my scenario since
Eq.(\ref{inequality}), realized as Eq.(\ref{inequality1}),
produces a terribly small monopolium which feels a negligible drag
force \cite{blanco}.

I have just shown that the proposed scenario is realized in this
naive scheme. I proceed next to investigate phenomenological
consistencies of the proposed picture.

\section{Monopolium abundance}

Monopolium has been associated with UHECR in various schemes
\cite{hill,sigl,blanco}. This association leads to a
phenomenological determination of its abundance. In here I look for
consistency between the phenomenological determined abundances and
the monopolium mean life obtained in my calculation.

What is the present number of monopolia? From the simple observation
that the mass density due to monopolia should not exceed the limit
on the mass density of the Universe imposed by Hubble's constant and
the deceleration parameter one gets, following
Preskill\cite{preskill}, for their density,

\bq \rho_{today} \sim 3. \; 10^{-18} \mbox{ monopolia/} cm^3 \;.
\eq

\noindent Is this number reasonable?

If I restrict the number of decays of monopolia to a few thousand
per year and per cubic parsec, a conservative estimate in agreement
with observational limits \cite{hill}, I get

\bq  \rho_{today} \sim  10^{-40} \mbox{ monopolia/} cm^3 \; ,
\label{jets}\eq

\noindent in agreement with the estimate of ref.\cite{blanco}. I
take the second estimate as more reasonable since it arises from
observation. A collateral result of the calculation is that
monopolia do not contribute greatly to the mass of the universe.

Is $\rho_{today}$ consistent with the used mean life?

In order to find out I have to calculate the number of monopolia at
formation. If I assume the standard scenario for nucleosynthesis
that requires that  monopolia do not dominate the mass of the
Universe at that time one gets \cite{preskill},

\bq \rho(kT = 1 \mbox{MeV})  \sim  5. \; 10^{13} \mbox{
monopolia/} cm^3 \; , \eq

\noindent which gives a density for today of

\bq \rho_{today} \sim  10^{-40} \mbox{monopolia/} cm^3 \;
.\label{meanlife}\eq
using $\rho(t) = \rho(0) e^{-\tau_U/\tau}$ for $\tau_U/\tau \sim 100
$.

\noindent Thus, within the validity of my scheme the two numbers,
Eqs. (\ref{jets}) and (\ref{meanlife}) are compatible.

A more sophisticated calculation following ref. \cite{preskill}
shows that monopole--anti-monopole annihilation is halted by the
expansion of the Universe. To achieve the same monopolia abundances,
a shorter mean life is necessary, which results in a better
verification of my fundamental Eq.(\ref{inequality}) and thus to a
more complete disappearance of free monopoles in the early Universe
in favor of monopolia.

I conclude from the above analysis that at present most monopolium
states are close to the annihilation levels, $n_f \sim 50$, and
therefore their binding energy is huge, at the level of

\bq E_b > 10^{14} \mbox{ GeV }  \; ,\eq

\noindent supporting Dirac's conjecture for the non observability of
monopoles.

\section{Monopolium detection}

My discussion would be semantic if the existence of  monopolium
could not be tested independently of the detection of monopoles. My
estimates for the density of  monopolia at present are similar to
those for which monopole detectors have been built. Moreover,
monopolium is easier to detect than monopoles themselves, because it
is a bound state and therefore one can use in addition spectroscopic
methods.

Cascading from large values of $n$ leads to a Larmor type spectrum

\bq \lambda_{radiation} \sim 16 \alpha ^2 \frac{h}{m c} n_n^3 \sim
32 \frac{\mbox{eV}}{m c^2} n^3\; \mbox{\AA} \; ,\eq

\noindent which for large values of $n$, where the formula is
certainly applicable, extends all the way into radio frequencies.
Experiments should look for a diffuse isotropic radio background
as a remnant from the nucleosynthesis period.

The core of galaxies, and of clusters of galaxies, provides us with
an environment of high density and energy where the existing
monopolia can be excited. One can therefore have a glimpse on the
high frequency part of the spectrum. Also,  monopolia may be broken
up there, and (anti)monopoles might be liberated, forming
immediately thereafter lightly bound  monopolia, which cascade down
slowly, by present time scales, to its annihilation regime showing
the low frequency part of the spectrum.

Monopolium can annihilate, at the level of a few thousand per year
and per cubic parsec, providing a huge amount of energy, $E
> 10^{15}$ GeV, in a small region of space-time which will produce UHECRs
in what Hill \cite{hill} calls a cataclysmic scenario, whose details
depend on the microscopic theory of monopole formation. Since the
mean life of monopolium is at least two orders of magnitude smaller
than the age of the Universe, these observations result from the
tale of the original population. Furthermore, showers of remnants
from the more numerous past annihilations should also be detectable.

Monopolium has properties, like magnetic polarizability and magnetic
multipole moments, which could be very interesting for laboratory
detection using the Lorentz and "dual" Lorentz force laws. Since
large electric fields are easier to construct than large magnetic
fields the dual Lorentz force might be instrumental for monopolium
detection in the laboratory.

\section{Conclusions}

The possibility of having (anti)monopoles in Nature is appealing. I
have presented a scenario for the Universe in which primordial
(anti)monopoles exist still today however, not as free particles,
but deeply bound in monopolium states. The crucial ingredient of my
proposal is that the mean free path of monopoles is much larger than
their capture radius and therefore they bind so tightly in
monopolium that they barely interact with the surrounding plasma,
surviving in this way the effect of the drag force.

Two distinctive quantities determine the consistency of the
various requirements, the temperature of monopolium formation ($kT
\sim 1$ MeV) and the mean life of the state ($\tau < 10^9$ years).
The outcome is an extremely massive particle protected from the
interaction with the medium in a strongly bound state, which
ultimately annihilates in a cataclysmic scenario giving rise to
UHECRs. The mean life of monopolium is at least two orders of
magnitude smaller than the age of the Universe, thus one is
contemplating only the decay of the tale of the original
population. Showers of remnants of the numerous past annihilations
should be also detectable.

The detection of monopolia, and therefore the existence of
monopoles, presents interesting signatures associated with the
monopolium spectrum, i.e., a diffuse isotropic radio frequency
background, localized (galaxies) high frequency excitation and
monopolium formation spectrum, and isotropically distributed very
energetic gamma rays associated with monopolium annihilation.

Finally, the electromagnetic properties of monopolium, associated
with the dual character of the electromagnetic interaction in the
presence of monopoles, are instrumental for designing new laboratory
experiments.

\section*{Acknowledgments}
I thank T. Sloan for a careful reading of the manuscript.
Discussions with Gabriela Barenboim, Jose Bordes, Carlos
Garc\'{\i}a-Canal, Huner Fanchiotti, Pedro Gonz\'alez, Santiago
Noguera and Arcadi Santamar\'{\i}a are acknowledged. This work was
supported by MCYT-FIS2004-05616-C02-01 and GV-GRUPOS03/094.

\end{document}